\documentclass[conference]{IEEEtran}
\IEEEoverridecommandlockouts
\usepackage{cite}
\usepackage{amsmath,amssymb,amsfonts}

\usepackage{algorithmic}
\usepackage{booktabs}
\usepackage{graphicx}
\usepackage{textcomp}
\usepackage{xcolor}
\usepackage{mathtools}
\usepackage[htt]{hyphenat}
\usepackage[hyphens]{url}
\PassOptionsToPackage{hyphens}{url}\usepackage{hyperref}
\def\BibTeX{{\rm B\kern-.05em{\sc i\kern-.025em b}\kern-.08em
    T\kern-.1667em\lower.7ex\hbox{E}\kern-.125emX}}
\begin{document}

\title{MORE: Measurement and Correlation Based Variational Quantum Circuit for Multi-classification\\
}

\author{\IEEEauthorblockN{Jindi Wu}
\IEEEauthorblockA{\textit{Department of Computer Science} \\
\textit{William \& Mary}\\
Williamsburg, VA, USA \\
jwu21@wm.edu}
\and
\IEEEauthorblockN{Tianjie Hu}
\IEEEauthorblockA{\textit{Department of Computer Science} \\
\textit{William \& Mary}\\
Williamsburg, VA, USA \\
thu04@wm.edu}
\and
\IEEEauthorblockN{Qun Li}
\IEEEauthorblockA{\textit{Department of Computer Science} \\
\textit{William \& Mary}\\
Williamsburg, VA, USA \\
liqun@cs.wm.edu}

}

\maketitle

\begin{abstract}
Quantum computing has shown considerable promise for compute-intensive tasks in recent years. For instance, classification tasks based on quantum neural networks (QNN) have garnered significant interest from researchers and have been evaluated in various scenarios. However, the majority of quantum classifiers are currently limited to binary classification tasks due to either constrained quantum computing resources or the need for intensive classical post-processing. In this paper, we propose an efficient quantum multi-classifier called \textbf{MORE}, which stands for \underline{m}easurement and c\underline{or}r\underline{e}lation based variational quantum multi-classifier. MORE adopts the \textit{same} variational ansatz as binary classifiers while performing multi-classification by fully utilizing the quantum information of a \textit{single} readout qubit. To extract the complete information from the readout qubit, we select three observables that form the basis of a two-dimensional Hilbert space. We then use the quantum state tomography technique to reconstruct the readout state from the measurement results. Afterward, we explore the correlation between classes to determine the quantum labels for classes using the variational quantum clustering approach. Next, quantum label-based supervised learning is performed to identify the mapping between the input data and their corresponding quantum labels. Finally, the predicted label is determined by its closest quantum label when using the classifier. We implement this approach using the Qiskit Python library and evaluate it through extensive experiments on both noise-free and noisy quantum systems. Our evaluation results demonstrate that MORE, despite using a simple ansatz and limited quantum resources, achieves advanced performance.
\end{abstract}


\section{introduction} \label{sec:intro}


Quantum computing derived from quantum mechanics can exponentially accelerate the solution of specific problems compared to classical computing, because of the unique properties of superposition and entanglement \cite{arute2019quantum}. Nevertheless, we are now in the Noisy Intermediate-Scale Quantum (NISQ) era, in which quantum machines contain just 50-100 qubits and are error-prone. Several months ago, IBM unveiled Osprey, the world's largest quantum computer with 433 qubits \cite{Collins2022ibm}. This effort will provide the groundwork for the next era of quantum-centric supercomputing. However, its computing procedure is still plagued with qubit decoherence errors, gate errors, and measurement errors \cite{clerk2010introduction}. Quantum Machine learning (QML) is one of the most attractive applications of quantum computing because of its demand for computing power and its resilience to noise \cite{biamonte2017quantum, schuld2015introduction, aimeur2006machine}. In recent years, many efforts have been made to develop QML techniques as the counterparts of classical ones, such as quantum k-nearest neighbor \cite{aimeur2006machine, dang2018image}, quantum support vector machines \cite{rebentrost2014quantum, li2015experimental, chatterjee2016generalized}, Quantum clustering \cite{maier2005quantum, aimeur2007quantum}, and quantum neural networks (QNNs) \cite{farhi2018classification, beer2020training, jeswal2019recent}. Since classical neural networks have achieved remarkable success in various fields \cite{shahin2001artificial, fu2022application, lakhan2022deep, elahi2022application, oza2022deep}, QNNs have also been attempted for data processing in many domains, including finance \cite{paquet2022quantumleap, el2018forecasting}, chemistry \cite{kandala2017hardware, sagingalieva2022hybrid} and healthcare \cite{hassan2020discrimination, esposito2022quantum, narain2016cardiovascular, mathur2021medical, umer2022integrated}.

\begin{figure}[t]
\centering 
\includegraphics[width=1\columnwidth]{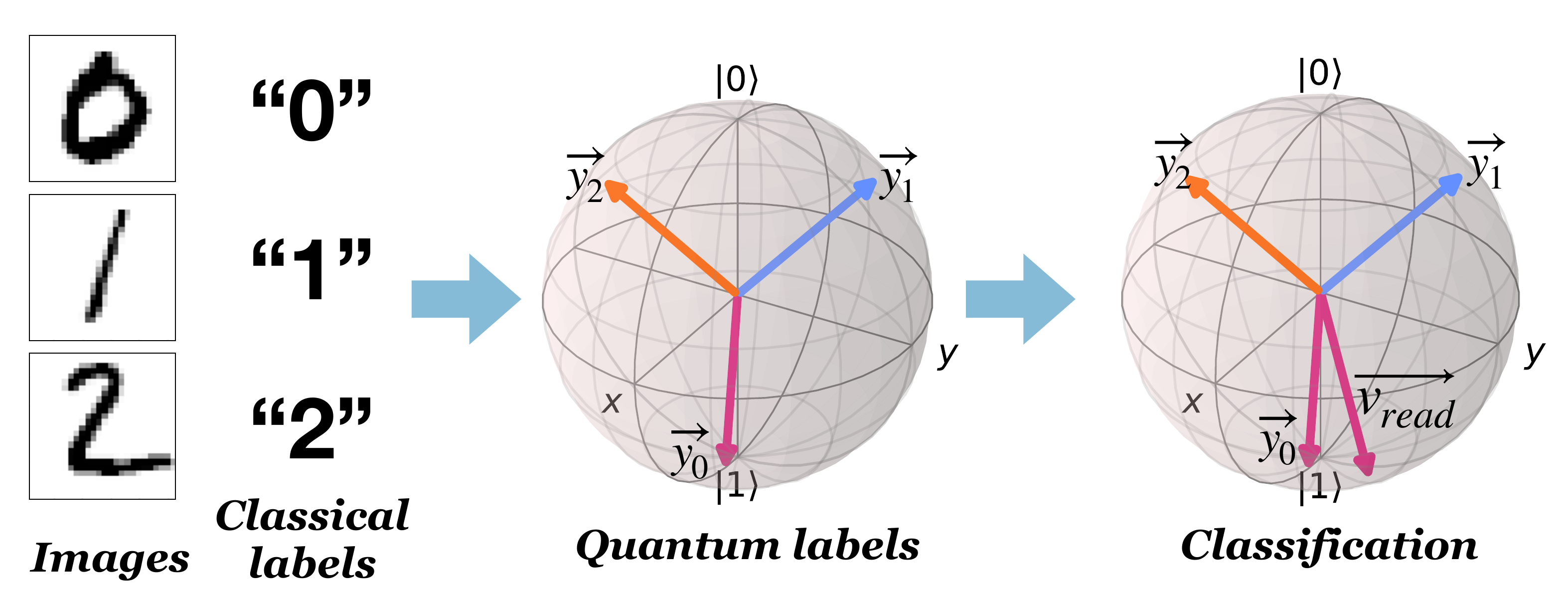}
\caption{Two-step MORE. First, the classical labels are converted into quantum labels in a two-dimensional Hilbert space based on the interclass correlation. Next, quantum label-based supervised learning is employed to minimize the difference between the readout state and its correct quantum label.}
\label{fig:twostep}
\end{figure}

Classification is a fundamental data processing technique in many domains, making QNN-based classifiers an area of significant interest. 
In 2018, \textit{Farhi} and \textit{Neven} proposed the first QNN classifiers for near-term processors \cite{farhi2018classification}. Following this, in 2019, \textit{Cong et al.} proposed quantum convolutional neural networks (QCNNs) for image processing, which adopt the architecture design of its classical analog \cite{cong2019quantum}. Based on these two quantum classifier designs, numerous variants are shown \cite{wu2022wpscalable, hur2022quantum, chalumuri2021hybrid, abohashima2020classification}. Moreover, some other QNN designs are suggested in \cite{henderson2020quanvolutional, blank2020quantum, schuld2017implementing, grant2018hierarchical, stein2022quclassi, abohashima2020classification}.

Prior works on QNN-based classification have predominantly focused on binary classification tasks, with relatively few addressing multi-classification problems. This is partly due to the fact that QNNs typically produce binary measurement results, meaning that a single readout qubit can only represent two classes. The popular solutions for QNN-based multi-classification are: (1) use extra ancillary qubits to represent labels for multiple classes. However, qubits are a limited resource on NISQ machines, and using more ancillary qubits reduces the number of qubits available for data processing, thus limiting the capacity of the QNN. (2) Select a subset of qubits at the end of the QNN circuit to serve as the readout for multi-class labels. However, if the one-hot encoding is used to represent labels, the number of classes a QNN-based classifier can handle is limited by the number of qubits available in the circuit. On the other hand, if binary encoding is used to represent labels, there may be cases where measurement results are not interpretable, which is a limitation that also applies to the previously mentioned solution. Specifically, the binary property of qubit measurement results enables an $n$-qubit state to include $2^n$ possible labels. Thus, given a $p$-classification task, where $p \in (2^{n-1}, 2^n)$, some output states will not map to any valid class. Especially when $p = 2^{n-1} + 1$, almost half of the label space is wasted. (3) Attaching classical fully-connected layers after QNNs is a common approach to mapping the QNN outcomes to multiple classes. However, this may incur high computational costs because the fully-connected layers contain a large number of parameters to train. Moreover, adding classical layers after QNNs may offset the potential quantum advantage of using QNNs for classification. And (4) decomposing the multi-classification task into several binary classification ones. The downsides include a complicated training process and lengthy training and inference times due to the need to train multiple binary classifiers. Therefore, there is a need for a novel QNN-based multi-classifier that features a concise variational quantum circuit, a streamlined classical post-processing procedure, and a straightforward training process.

In this article, we propose \textit{MORE} approach, a novel QNN multi-classifier that utilizes measurements and interclass correlations. MORE is designed to be resource-efficient, featuring a simple ansatz similar to that used in binary classification, with only one readout qubit at the end of the circuit. MORE is a two-step approach, as shown in fig.~\ref{fig:twostep}. First, it converts the classical labels of training data into quantum labels, which are quantum states reconstructed from the measurement results of the readout qubit. The quantum label of each class is determined using the variational quantum clustering method that considers the correlation between classes. Then, quantum label-based supervised learning is conducted to converge the QNN and achieve the desired performance quickly. Compared to prior QNN-based multi-classifiers, MORE requires fewer qubits and quantum gates, resulting in fewer gate and decoherence errors during the quantum state evolution. Moreover, MORE is computationally efficient, with a fixed amount of quantum computation and linearly increasing classical computation, allowing scalability to more complex tasks.

The contribution of this work is fourfold:
\begin{itemize}
    \item We present the attempt to design efficient quantum multi-classifiers by leveraging the full quantum information contained in a single qubit.
    \item We offer a technique for converting the classical labels of training data into quantum states by investigating interclass correlations.
    \item We introduce a quantum label-based quantum supervised learning approach that includes a loss adjuster to improve the quality of the QNN model.
    \item We implement the proposed MORE approach and evaluate it from many perspectives using comprehensive experiments. And our experiment results demonstrate the advantages of our proposed approach.
\end{itemize}

In the remaining content of this article, we provide some basic knowledge needed for understanding this article in section~\ref{sec:pre}. We then introduce our motivation for the proposed approach by discussing quantum state tomography, which is a method for reconstructing quantum states, in Section~\ref{sec:tomo}. The proposed two-step approach MORE is detailed in section~\ref{sec:method}. The first step, variational quantum clustering based on interclass correlation, is described in subsection~\ref{sec:clt}, then the second step, quantum label-based supervised learning, is discussed in subsection~\ref{sec:super}. Then, we present the evaluation setup and results in Section~\ref{sec:eval}, followed by a review of related works in Section~\ref{sec:related}. Finally, we conclude our proposed MORE approach in Section~\ref{sec:conclusion}.








\section{Preliminaries} \label{sec:pre}

\subsection{Quantum state and visualization}

A qubit (short for a quantum bit) is the information carrier in the quantum computing/communication channel \cite{kilin2001quantum}. A qubit is defined as a two-dimensional Hilbert space with two orthonormal bases $|0\rangle$ and $|1\rangle$, which are known as computational bases in two-level quantum computing. These computational bases are usually represented as vectors $|0\rangle = [1, 0] ^\top$ and $|1\rangle = [0,1]^\top$. Due to the unique qubit characteristic of \textit{superposition}, the state of a qubit can be represented as the sum of two computational bases weighted by (complex) amplitudes:
\begin{align}
    |\psi \rangle = \alpha |0\rangle + \beta |1\rangle&= \begin{bmatrix}
            \alpha \\
           \beta 
         \end{bmatrix}
\end{align}
where $\alpha$ and $\beta  \in \mathbb{C}$, and $|\alpha| ^ 2 + |\beta|^2 = 1$. $|\alpha| ^ 2$ and $|\beta|^2$ are the probability of obtaining states $|0\rangle$ and $|1\rangle$ after multiple measurements, respectively. 

\begin{figure}[t]
\centering 
\includegraphics[width=0.6\columnwidth]{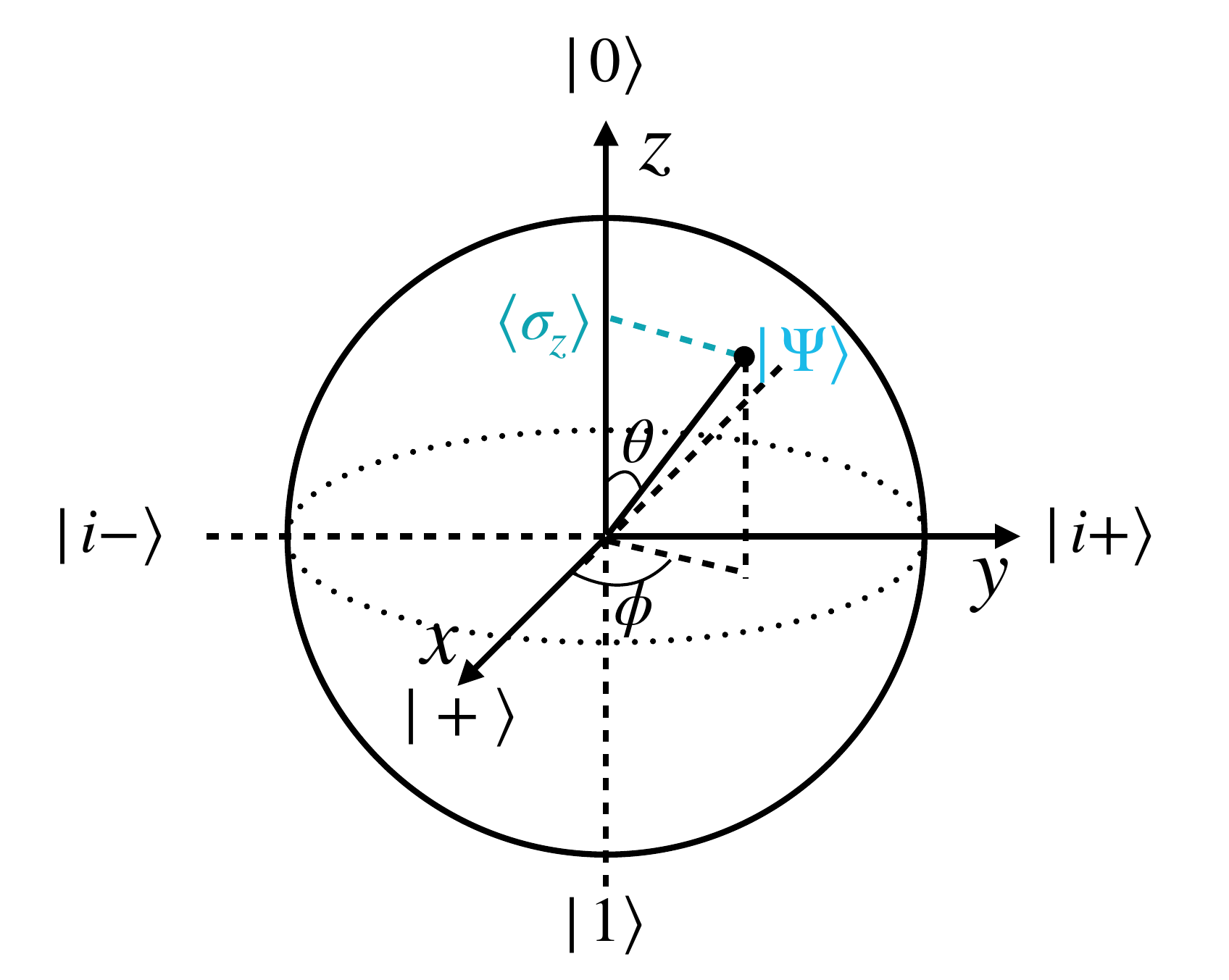}
\caption{Bloch sphere and z-measurement}
\label{fig:block}
\end{figure}

The Bloch sphere is a valuable tool for visualizing the state of a single qubit, as shown in Fig.~\ref{fig:block}. It encompasses all possible states of a qubit, making it an excellent representation of a two-dimensional Hilbert space. In this article, we will use the Bloch sphere to illustrate our proposed method clearly. Every pure state of a qubit can be mapped to a distinct point on the surface of the Bloch sphere, whereas mixed states correspond to points within the sphere. The state of a qubit on the Block sphere can be described with two real parameters, $\theta$ and $\phi$,
\begin{align}
    |\psi \rangle = \cos{\frac{\theta}{2}} |1\rangle + e^{i \phi} \sin{\frac{\theta}{2}} |1\rangle
\end{align}
where $\theta \in [0, \pi]$ and $\phi \in [0, 2\pi]$. I.e., $\theta=0$ for $|0\rangle$ and $\theta=\pi$ for $|1\rangle$, and global phase $\phi$ can be any value.

\subsection{Quantum measurement}
Quantum measurement is the retrieval of the numerical information stored in a qubit. A measurement result is +1 for state $|0\rangle$ and -1 for state $|1\rangle$ according to a specified probability distribution associated with the quantum state. Therefore, numerous measurements are required to determine the exact quantum state. The final result of the quantum measurement is the expected value of all outcomes.

Observables are used to understand the properties of a quantum system and can be measured. Mathematically, observables are formulated as Hermitian operators that map Hilbert space onto themselves. For a valid observable, its eigenvalues are real numbers and can be the outcomes of measurement. Moreover, observables can form an orthonormal basis of the target Hilbert space, which will be the state of the quantum system after measurement. The observables considered in this article are Pauli matrices:
\begin{equation}
    \sigma_x= \left[ \begin{array}{ccc}
    0 & 1\\
    1 & 0\end{array} \right],
    \sigma_y= \left[ \begin{array}{ccc}
    0 & -i\\
    i & 0\end{array} \right],
    \sigma_z = \left[ \begin{array}{ccc}
    1 & 0\\
    0 & -1\end{array} \right].
\end{equation}
These Pauli matrices span a complex two-dimensional Hilbert space (a qubit). The projection measurement is to extract quantum information by operating on the interested observable and the density matrix of the target quantum state
\begin{equation}
    \langle \sigma \rangle = Tr(\sigma |\psi\rangle \langle \psi|)
\end{equation}
where $|\psi\rangle \langle \psi|$ generates the density matrix.
For the general z-measurement, the state vector is projected onto the z-axis of the Bloch sphere, and the corresponding value on the z-axis is the expectation of the measurement results, as shown in Fig.~\ref{fig:block}.

\subsection{Variational quantum algorithm}

The variational quantum algorithm (VQA) is the standard approach to performing QNN. It processes prepared quantum information by applying a series of parametric quantum gates, ultimately producing an output through measurement. As an example, a binary quantum classifier is implemented using VQA in \cite{farhi2018classification}. It has an ($n$+1)-qubit circuit, where the first $n$ qubits are prepared using an encoding method (such as angle, basis, or amplitude encoding) to represent specific information. The final qubit acted as a readout, generating the output through measurements. These qubits then pass a sequence of quantum gates with trainable parameters $U(\theta) = \prod_{l=1}^N U_l(\theta_l)$, where $\theta$ is a set of parameters.  A measurement outcome of 1 corresponds to one class, while a result of -1 corresponds to the other class. 
VQA uses a hybrid quantum-classical procedure to iteratively optimize the trainable parameters.
The popular optimization approach includes gradient descent \cite{sweke2020stochastic}, parameter shift \cite{wierichs2022general}, and gradient-free techniques, such as COBYLA. All of the methods take the training data as input and evaluate the model performance by comparing the generated and correct labels. Based on this evaluation, the methods update the model parameters for the next round, repeating the process until the model converges and achieves the desired performance. The hybrid method performs the evaluation and parameter selection on a classical machine, while the model inference is carried out on a quantum machine.








\section{Quantum state tomography} \label{sec:tomo}

Quantum state tomography (QST) is a technique to reconstruct an unknown quantum state using its measurement results \cite{toninelli2019concepts}. The measured observables must form a basis in the Hilbert space so that all state information can be recorded and used to recover the state. In the present age of NISQ, when the number of qubits is limited, QST is an essential method for retrieving the complete information stored in a quantum system. Nevertheless, as the number of qubits grows, the number of measurements needed and the complexity of state reconstruction increase exponentially. Hence, for the sake of simplicity, we consider the reconstruction of a single qubit using three observables to restore its state in a two-dimensional Hilbert space in this work.

\begin{figure*}[t]
\centering 
\includegraphics[width=1\linewidth]{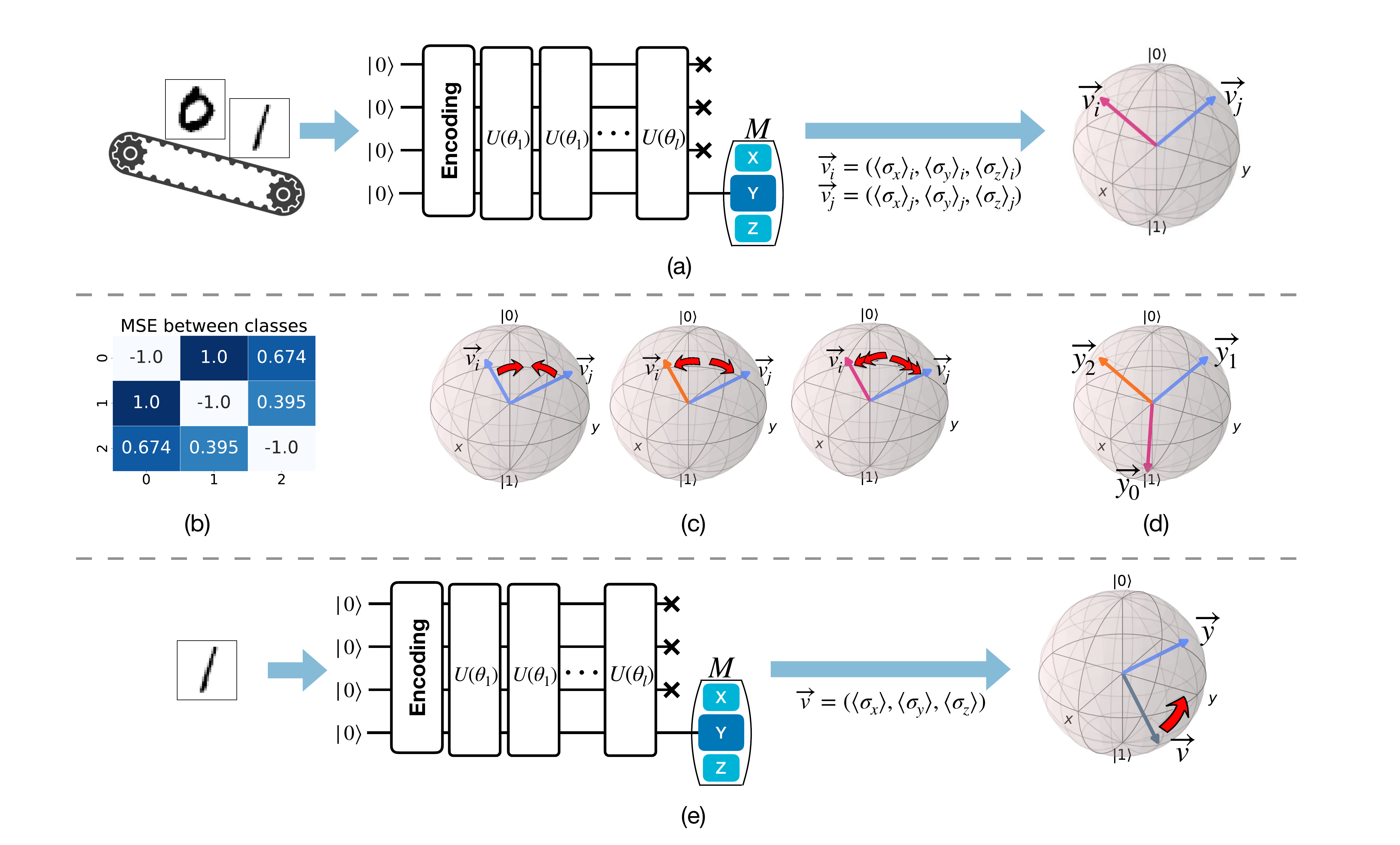}
\caption{MORE overview. Step 1: variational quantum clustering (a)-(d). (a) The QNN takes a pair of training instances as input and reconstructs their single-qubit readout states using x-, y-, and z-measurements. (c) QNN's parameters are tuned to adjust the distance between readout states depending on (b) the interclass correlations represented by MSE. (d) After iterative training, the distribution of readout states in Hilbert space corresponds to that of classes in feature space. The centroid of readout states within the same class is the quantum label for this class. Step 2: quantum label-based supervised training (e). The readout state tends to approach its corresponding quantum label during the training process. During inference, the predicted label of a test instance is the quantum label closest to its readout state.}
\label{fig:overview}
\end{figure*}

Any arbitrary density matrix of a 1-qubit state can be expressed as a linear combination of Pauli matrices (basis of two-dimensional Hilbert space) as
\begin{equation}
\begin{split}
    \rho &= \frac{1}{2}(I + r_x \sigma_x + r_y \sigma_y + r_z \sigma_z)\\
    &= \frac{1}{2}\left[ \begin{array}{ccc}
    r_z + 1 & r_x - r_y\\
    r_x + r_y & -r_z + 1\end{array} \right]
\end{split}
\end{equation}
where $r$ denotes real number and $r_x^2 + r_y^2 + r_z^2 = 1$. For the typical qubit measurement using observable $\sigma_z$, as used by most quantum applications, the expectation of the measurement result is 
\begin{equation} 
\label{eq:obz}
 \langle \sigma_z \rangle = Tr\bigg(\frac{1}{2}(I + r_x \sigma_x + r_y \sigma_y + r_z \sigma_z) \sigma_z \bigg) = r_z
\end{equation}
Eq.~\ref{eq:obz} demonstrates that the expectation of measurement results is directly related to the density matrix $\rho$ of interest. In this case, however, only diagonal entries of $\rho$ can be retrieved, while some useful information remains untouched. Hence, in order to rebuild the density matrix $\rho$ completely, the measurements on observables $\sigma_x$ and $\sigma_y$ are necessary to obtain $r_x$ and $r_y$. 
 
From the perspective of the geometric representation, the expectation values $\langle \sigma_x \rangle$, $\langle \sigma_y \rangle$ and $\langle \sigma_z \rangle$ are exactly the projections of the state vector on the x-, y- and z-axis of the Bloch sphere, respectively. Therefore, after getting sufficient measurement results, we can restore the quantum state in the Bloch sphere for an intuitive interpretation. In other words, the interested 1-qubit state is a state vector with an unknown direction in the Bloch sphere. By projecting the vector on the three axes, we can figure out the direction of the state vector, which contains all information about the state.

\section{Measurement and correlation-based QNN multi-classifier} \label{sec:method}

In this section, we present MORE, an efficient quantum multi-classifier that employs a simple QNN with a single readout qubit. We use QST to reconstruct the state of the readout qubit to retrieve the entire information from it. This allows us to fully leverage information of a single qubit to address complex multi-classification tasks via two steps. First, we employ the variational quantum clustering method to convert the classical labels of the dataset into 1-qubit states, taking into account interclass correlations. Next, we proceed to train the QNN using a quantum label-based supervised learning approach, where the QNN learns to map an input to its corresponding quantum label. After completing the training process, the model can accurately categorize new data by comparing the readout state of the data with quantum labels and assigning it to the appropriate class based on the shortest distance.






\subsection{Overview}
The overview of MORE is presented in Fig.~\ref{fig:overview}. The goal is to solve a multi-classification task that assigns appropriate labels to classical input data. A training dataset consisting of $K$ subsets is provided, $D = \{D_1, D_2, \dots , D_K\}$. The subset $D_k$ is a collection of training data in class $k$, with each training data instance being a pair of data point $x$ and its correct classical label $y$. MORE addresses this problem by utilizing a QNN to learn the pattern that characterizes the mapping between the input data point and its correct label from the training dataset. The learning of MORE is a two-step procedure. First, the classical labels $y$ are converted into quantum labels $\overrightarrow{y}$ in the two-dimensional Hilbert space. This step employs the quantum clustering method while considering the inherent correlation between classes. (Fig.~\ref{fig:overview}(a-d) and Section~\ref{sec:clt}). Then, quantum label-based supervised learning is undertaken to fine-tune the parameters (Fig.~\ref{fig:overview}(e) and Section~\ref{sec:super}).


The ansatz of MORE is shown in Fig~\ref{fig:overview}(a). Its circuit includes three layers: (1) Encoding layer: the encoding layer quantizes the classical input data by preparing the quantum states based on classical data values. MORE is compatible with any general encoding scheme, such as angle and amplitude encoding. (2) Variational data processing layer $U(\theta)$: this layer transforms the prepared state using its parameterized gates. The parameter set $\theta$ of the layer is iteratively adjusted based on MORE's performance using a hybrid classical-quantum method, as introduced in section~\ref{sec:pre}. The optimal parameters should enable MORE to assign a label $y$ to unseen input $x$ with high accuracy. (3) Measurement layer $M$: One qubit is chosen as the readout in this layer, and its observable $\sigma_x$, $\sigma_y$, and $\sigma_z$ are measured to produce a three-dimensional vector (reconstructed state vector $\overrightarrow{v_i}$ and $\overrightarrow{v_j}$).

\begin{figure}[h]
\centering 
\includegraphics[width=0.8\columnwidth]{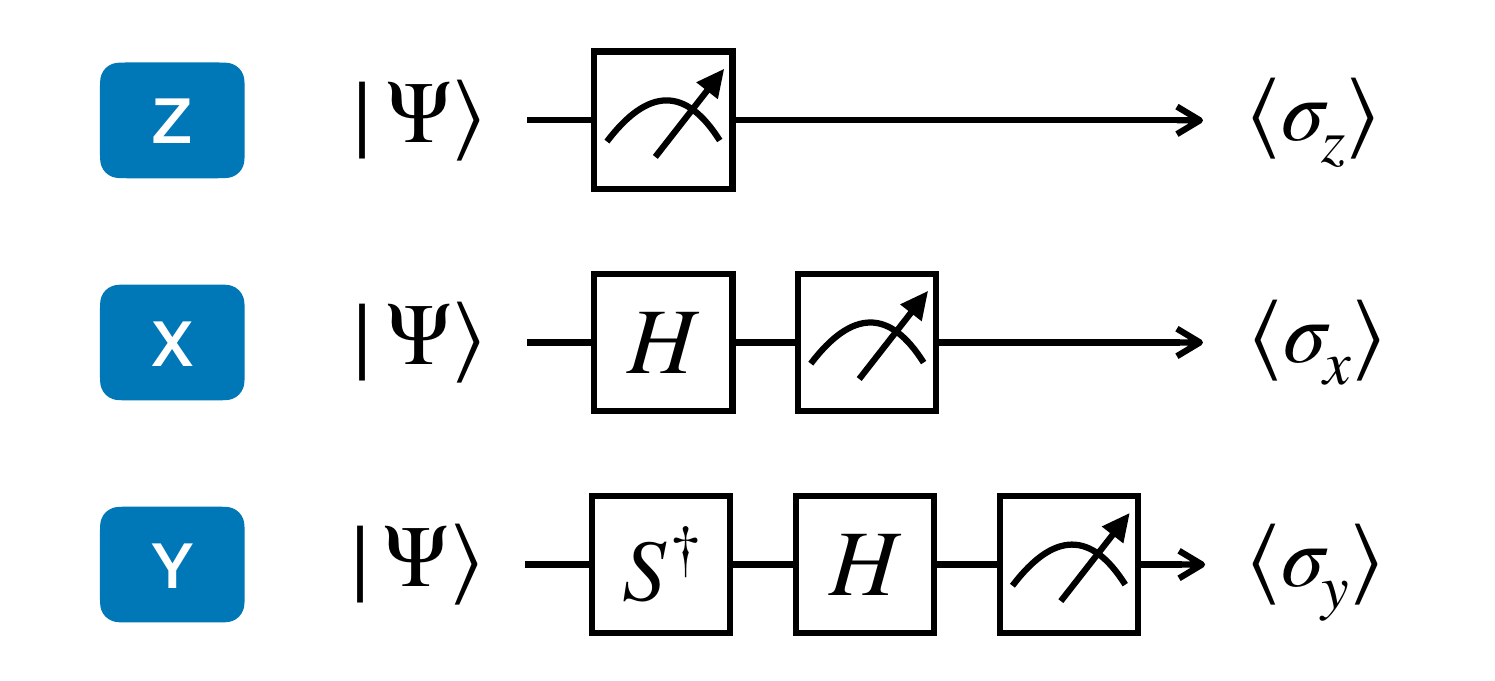}
\caption{Measurement unit of MORE}
\label{fig:munit}
\end{figure}

As the quantum measurement 
operator in superconducting hardware implementation can only measure the state on the z-axis, MORE requires two additional measurement units to measure the observables $\sigma_x$ and $\sigma_y$, as shown in Fig.~\ref{fig:munit}. The unit for observable $\sigma_x$ inserts an $H$ gate before the measurement operator, which can be conceptualized as rotating the x-axis of the Bloch sphere to the location of the z-axis first and then projecting the state vector to the new z-axis. Similarly, for observable $\sigma_y$ the readout qubit passes an $S^\dagger$ gate and an $H$ gate, followed by the measurement operator in the measurement unit. In the measurement layer of the QNN, these three measurement units are applied in succession to construct the state vector $\overrightarrow{v} = (\langle \sigma_x \rangle, \langle \sigma_y \rangle, \langle \sigma_z \rangle)$.

\subsection{Class correlation-based variational quantum clustering} \label{sec:clt}

To perform multi-classification on the 1-qubit state vectors, it is necessary to determine the distribution of classes in a two-dimensional Hilbert space first. This involves converting classical labels to quantum labels. To accomplish this, we use the \textit{variational quantum clustering} method while investigating the correlation between classes. Deep clustering is a technique in classical machine learning that concurrently learns the parameters of neural networks and the cluster assignments for processed inputs \cite{caron2018deep}. Here, we borrow this idea and apply it to variational quantum clustering. Specifically, we train the parameters of the QNN to map readout state vectors from the same class to a particular quantum state.

\textbf{Class correlation.}
First of all, we calculate the correlation between classes by measuring the difference between the data instances belonging to different class collections. To be specific, we first pre-process the training data using Principal Component Analysis (PCA) to extract critical features and downscale the data to a size that is suitable for our quantum circuit. Then, the average pattern set $D_{pattern}$ is created by averaging the data instances of the class dataset $D_k$,
\begin{equation}
    D_{pattern} = \{\overline{x}_1, \overline{x}_2, \dots, \overline{x}_K\}
\end{equation}
where $\overline{x}_k = \frac{1}{|D_k|} \sum_{i=1}^{|D_k|} x^i_k$, $k \in [1, K]$. The correlation between the average pattern of classes is then recorded in a $K \times K$ array, known as scaler array $S$. For example, we use Mean square error (MSE) to quantify the difference between two average patterns. As a result, the entry $(i, j)$ of the scaler array $S$ can be calculated by 
\begin{equation}
    S_{i, j} = MSE(\overline{x}_i, \overline{x}_j)
\end{equation}
The larger the MSE value, the lower the correlation. The original diagonal entries of $S$ are 0s, indicating that the correlation between two identical patterns is maximal. We modify these 0s to negative values, which will be explained and utilized later in this section. Fig.~\ref{fig:overview}(b) shows an example of the scaler array $S$ that stores the correlations between classes `0', `1', and `2' from the MNIST handwritten digit dataset.

\textbf{Clustering dataset.}
We then prepare the clustering dataset by pairing the instances in the given dataset $D$. The clustering dataset consisting of data pairs: $D_{cluster} = \{ [(x_i, y_i), (x_j, y_j)]_t\}_{t=1}^{N}$ where $i, j \in [1, K ]$ and $N$ is a small number to ensure the model can fast converge. In our evaluation, we set it to $\binom{5K}{2}$. Each data pair is independently sampled from $D$ across all classes, i.e., $(x_i, y_i)$ and $(x_j, y_j)$ are two sampled instances from class $i$ and $j$.


\textbf{Clustering loss.}
In a training step, the QNN $U(\theta)$ sequentially takes as input of a data pair $x_i$ and $x_j$ and generates two three-dimensional state vectors $\overrightarrow{v_i} = (\langle \sigma_x \rangle_i, \langle \sigma_y \rangle_i, \langle \sigma_z \rangle_i)$ and $\overrightarrow{v_j} = (\langle \sigma_x \rangle_j, \langle \sigma_y \rangle_j, \langle \sigma_z \rangle_j)$. We express the loss function as
\begin{equation}
    \mathcal{L}_{clt} = - S_{i,j} \times \mathcal{D}ist (\overrightarrow{v_i}, \overrightarrow{v_j})
\end{equation}
where
\begin{equation}
    \mathcal{D}ist(\overrightarrow{v_i}, \overrightarrow{v_j}) = 1 - \frac{\overrightarrow{v_i} \cdot \overrightarrow{v_j}}{|| \overrightarrow{v_i}|| \ ||\overrightarrow{v_j} ||}
\end{equation}
The function  $\mathcal{D}ist(\overrightarrow{v_i}, \overrightarrow{v_j})$ computes the Cosine distance between two vectors $\overrightarrow{v_i}$ and $\overrightarrow{v_j}$, and the coefficient $S_{i,j}$ serves as the scaler of cosine distance. We assume that the distribution of state vectors in a two-dimensional Hilbert space is relevant to the distribution of data in feature space. Therefore, the clustering step aims to bring the generated state vectors $\overrightarrow{v_i}$ and $\overrightarrow{v_j}$ closer together if $x_i$ and $x_j$ belong to the same class, and farther apart otherwise. In addition, the distance between $\overrightarrow{v_i}$ and $\overrightarrow{v_j}$ will be larger if their class correlation is significantly smaller. Specifically, as illustrated in Fig.~\ref{fig:overview}(c), if $x_i$ and $x_j$ from the same class, the negative scaler value $S_{i, j}$ where $i=j$, will minimize the cosine distance between $\overrightarrow{v_i}$ and $\overrightarrow{v_j}$. Otherwise, the distance between $\overrightarrow{v_i}$ and $\overrightarrow{v_j}$ will be maximized, and the final distance between them depends on the value of $S_{i, j}$. The higher the correlation, the closer they will be.

\textbf{Quantum labels.}
After the training of quantum clustering, the state vectors of data instances belonging to the same class are oriented in similar directions (a cluster). The centroid of these directions is recorded as the quantum label of the class. So far, we have mapped the classcial labels $y$ into quantum labels $\overrightarrow{y}$, which are distributed in the Hilbert space based on the correlation between classes, as shown in Fig.~\ref{fig:overview}(d). Then we substitute the classical labels $y$ in the training dataset $D$ with quantum labels $\overrightarrow{y}$, and now we are ready to proceed with quantum label-based quantum supervised learning.

\subsection{Quantum label-based supervised learning} \label{sec:super}

After the clustering step, we initially train the parameters $\theta$ of QNN $U(\theta)$ to determine the quantum labels for classes. However, as the $U(\theta)$ is trained on a small subset of the training data, it may not be optimal for classification. Thus, to enhance the performance of $U(\theta)$, we fine-tune the parameters by conducting quantum label-based supervised learning. We use the dataset containing the training data points $x$ and their quantum labels $\overrightarrow{y}$ to train $U(\theta)$. The goal is to map the readout state vector of inputs to be as close as possible to their corresponding quantum labels, as shown in Fig.~\ref{fig:overview} (e). In this step, the $U(\theta)$ takes one training instance as input and generates the 1-qubit state vector $\overrightarrow{v}$. Then, we compare the generated vector with its corresponding quantum label $\overrightarrow{y}$, and express the loss function of quantum label-based supervised learning as
\begin{equation}
\label{eq:sup_loss1}
    \mathcal{L}_{sup} = \mathcal{D}ist(\overrightarrow{v}, \overrightarrow{y})
\end{equation}
The training procedure aims to minimize the value of $ \mathcal{L}_{sup}$. For inference, the class of the unseen data is determined by
\begin{equation}
    y = argmin_k \{\mathcal{D}ist(\overrightarrow{v}, \overrightarrow{y_k}) \}
\end{equation}
where $\overrightarrow{y_k}$ is the quantum label for class $k \in [1, K]$.

\begin{figure}[t]
\centering 
\includegraphics[width=0.9\columnwidth]{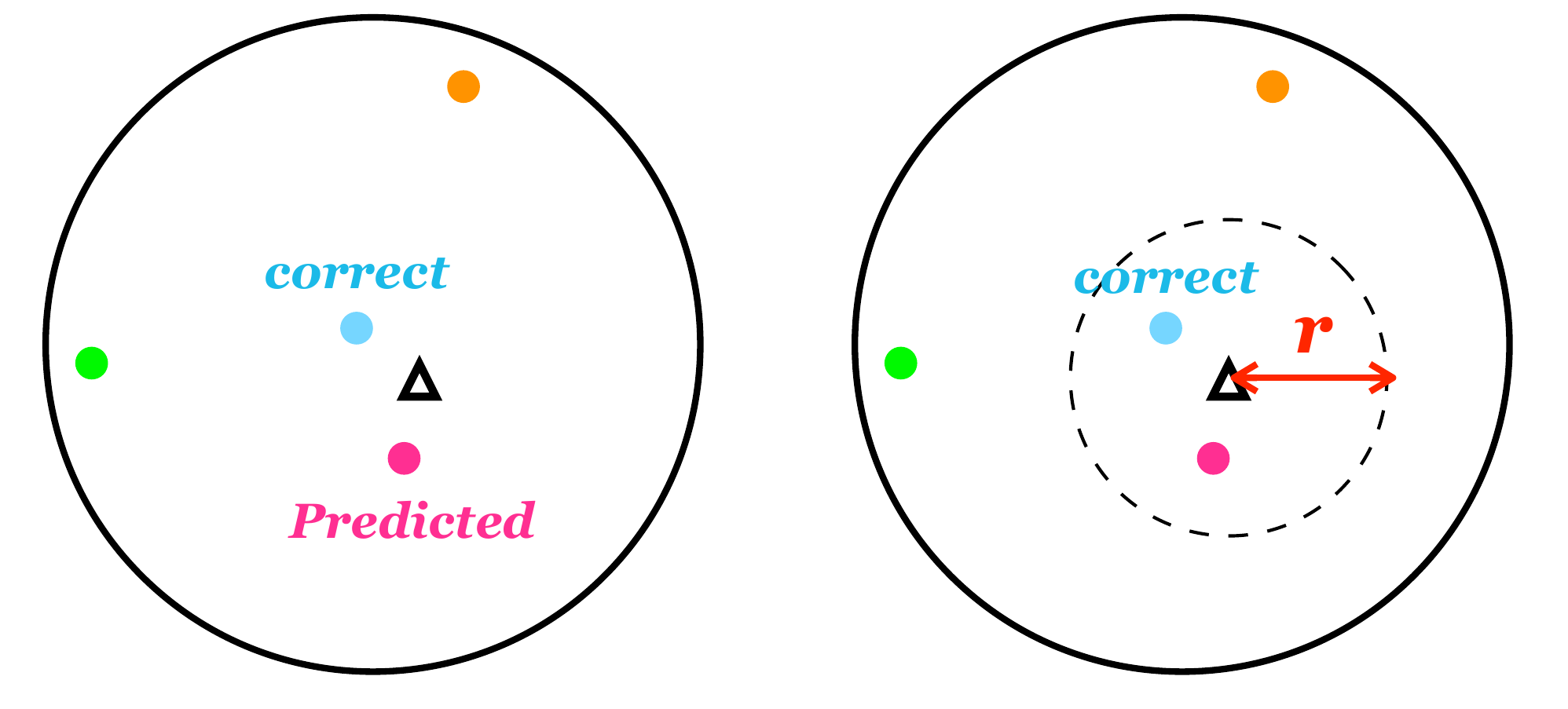}
\caption{Curse of density (left) and loss adjuster (right)}
\label{fig:lossreg}
\end{figure}

However, misclassifications are more likely to occur when there are a large number of classes and their quantum labels are closely distributed in the Hilbert space, which is known as the curse of density (see Fig.~\ref{fig:lossreg} left). The circle is a side view of the Bloch sphere, and the colored dots are quantum labels. The triangle represents the readout state of an input. It may be classified into a nearby but incorrect class (represented by the pink dot) due to the small distance between the readout state and the pink quantum label. To improve the accuracy of the classifier in scenarios with crowded quantum labels, we introduce a loss adjuster $\mathcal{R}$.

In each training step, if the distance between the readout state and its correct quantum label is less than a threshold $r$, we also consider the other quantum labels whose distance from the readout state is less than the threshold, as shown in Fig.~\ref{fig:lossreg} right. Specifically, when $\mathcal{D}ist(\overrightarrow{y}, \overrightarrow{v}) \leq r$, we define the loss regulator as
\begin{equation}
\label{eq:reg}
    \mathcal{R} = \sum_{k \in K'} \mathcal{D}ist(\overrightarrow{y_k}, \overrightarrow{v})
\end{equation}
where $K'$ is a subset of classes whose quantum label $\overrightarrow{y_k}$ satisfies $\mathcal{D}ist(\overrightarrow{y_k}, \overrightarrow{v}) \leq r$.

We then assign a weight factor $w \in [0, 1]$ to the supervised loss term. Hence, the overall objective function of the quantum labels-based training process is 
\begin{equation}
\label{eq:loss_r}
    \mathcal{L}_{sup+\mathcal{R}} = w \mathcal{L}_{sup} - (1-w)\mathcal{R}.
\end{equation}
By minimizing the value of $\mathcal{L}_{sup+\mathcal{R}}$, this objective function tends to bring the readout state closer to its correct quantum label while moving farther away from the incorrect quantum labels surrounding it, resulting in the readout state being closest to its correct label among crowded quantum labels. The selection strategy for the values of $r$ and $w$ will be discussed in Sec.~\ref{sec:reg}.






\section{Evaluation} \label{sec:eval}
We conducted empirical studies in various scenarios to evaluate the performance of the proposed quantum multi-classifier MORE. To implement MORE and related baselines, we used Python 3.8 and the IBM Qiskit package \cite{qiskit} to simulate quantum systems both with and without noise. For simulating noisy systems, we employed several noisy backends, including \texttt{FakeAuckland}, \texttt{FakeAthensV2}, and \texttt{FakeBelemV2}. The source code used to generate the experiment results is available in \href{https://github.com/Jindi0/MORE.git}{github.com/Jindi0/MORE}. 

\begin{figure*}[t]
\centering 
\includegraphics[width=1\linewidth]{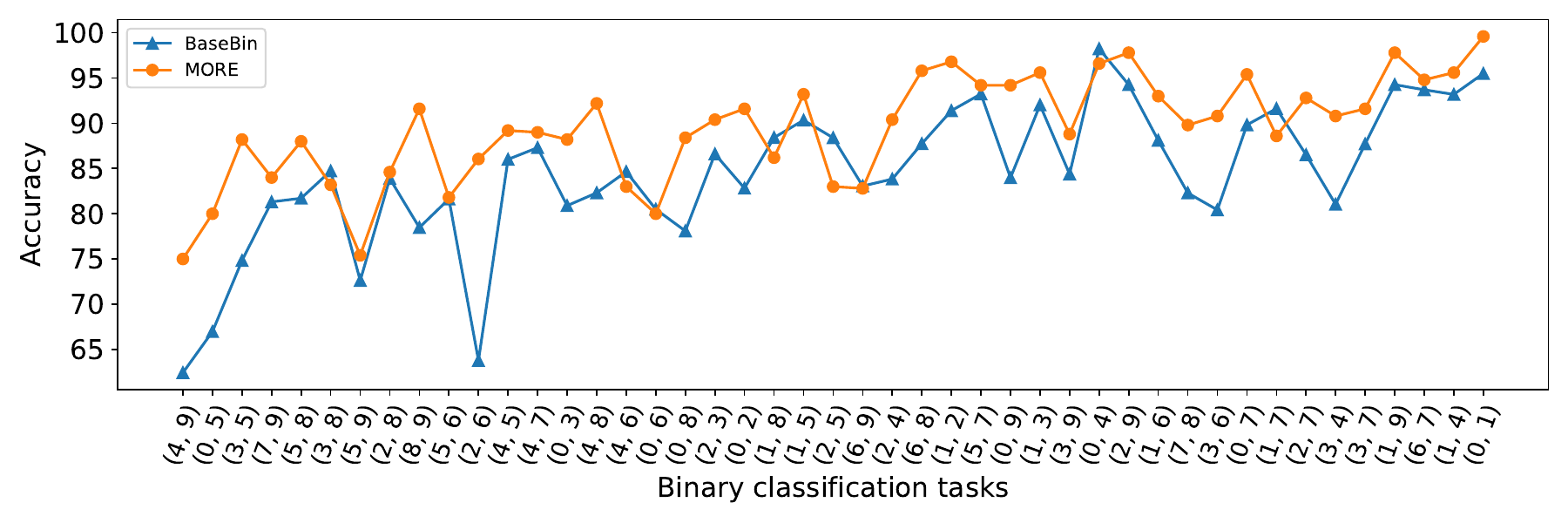}
\caption{Accuracy of binary classification tasks}
\label{fig:eval-bin}
\end{figure*}

\begin{figure*}[t]
\centering 
\includegraphics[width=1\linewidth]{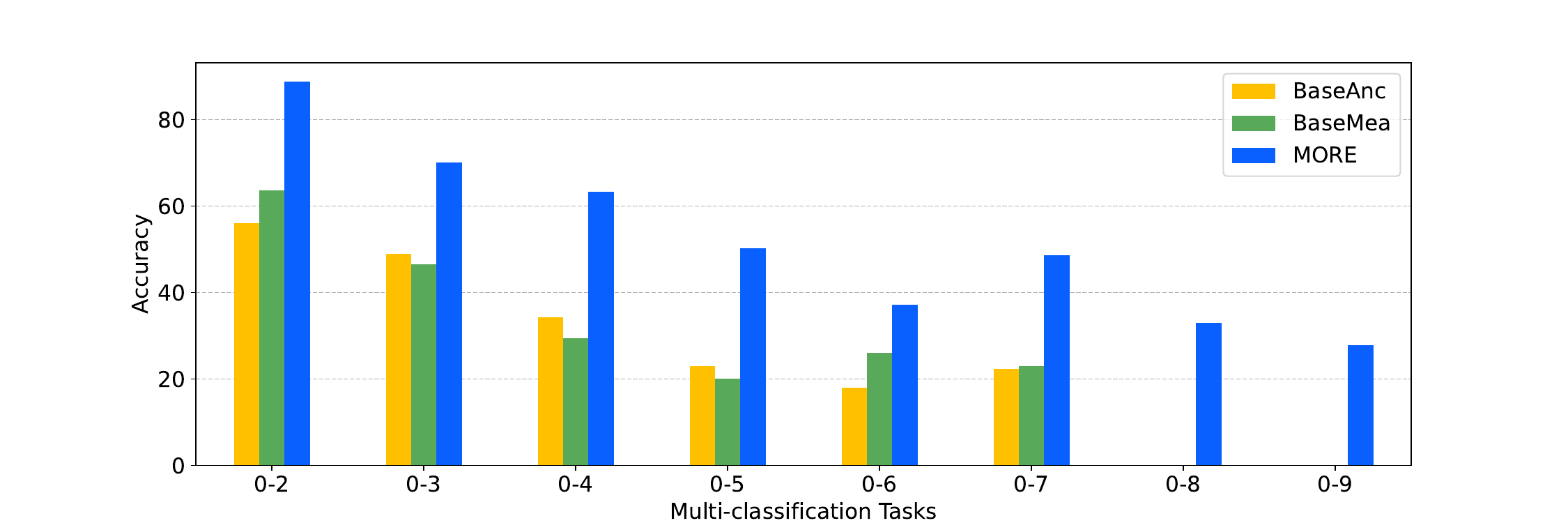}
\caption{Accuracy of multi-classification tasks}
\label{fig:eval-mul}
\end{figure*}

\textbf{Dataset:} (1)We utilize the \textbf{MNIST} dataset \cite{deng2012mnist} to conduct experiments in the noiseless quantum system. The MNIST is a widely used benchmark for image classification tasks, consisting of ten classes of hand-written digits ranging from 0 to 9. We randomly select 1,000 training and 200 test images from each class and reduce their dimensions to 8 using PCA. (2) Additionally, we evaluate MORE using the \textbf{Iris} dataset \cite{Iris} in simulated noisy quantum systems. The Iris dataset contains 150 instances classified into three distinct classes, each consisting of 50 instances with 4 pixels. We use 70\% of the instances (105 instances) for training, while the remaining 30\% (45 instances) are reserved for testing purposes.

\textbf{Model implementation:}
We develop the QNN classifiers based on Qiskit \texttt{NeuralNetworkClassifier} class, and use the optimizer \texttt{COBYLA} to update trainable parameters. MORE employs an 8-qubit QNN for the MNIST dataset and a 4-qubit QNN for the Iris dataset, with 91 and 39 trainable parameters, respectively. We build QNNs based on the design principles for quantum convolutional neural networks proposed in \cite{cong2019quantum}. For the readout, we choose the last active qubits at the end of the circuit and measure it with $\sigma_x$, $\sigma_y$, and $\sigma_z$ observables. To efficiently convert classical labels to quantum labels during the clustering step, we randomly select five instances from each class and form pairs from the resulting dataset. The clustering dataset consists of $\binom{5K}{2}$ pairs in total, where $K$ is the number of classes. We use MSE to calculate the interclass correlations. And to make the quantum labels spread out as much as possible, the MSE between different classes is normalized to the interval [0.5, 1]. During supervised learning, then, the entire training dataset is used.


\begin{figure*}[t]
\centering 
\includegraphics[width=1\linewidth]{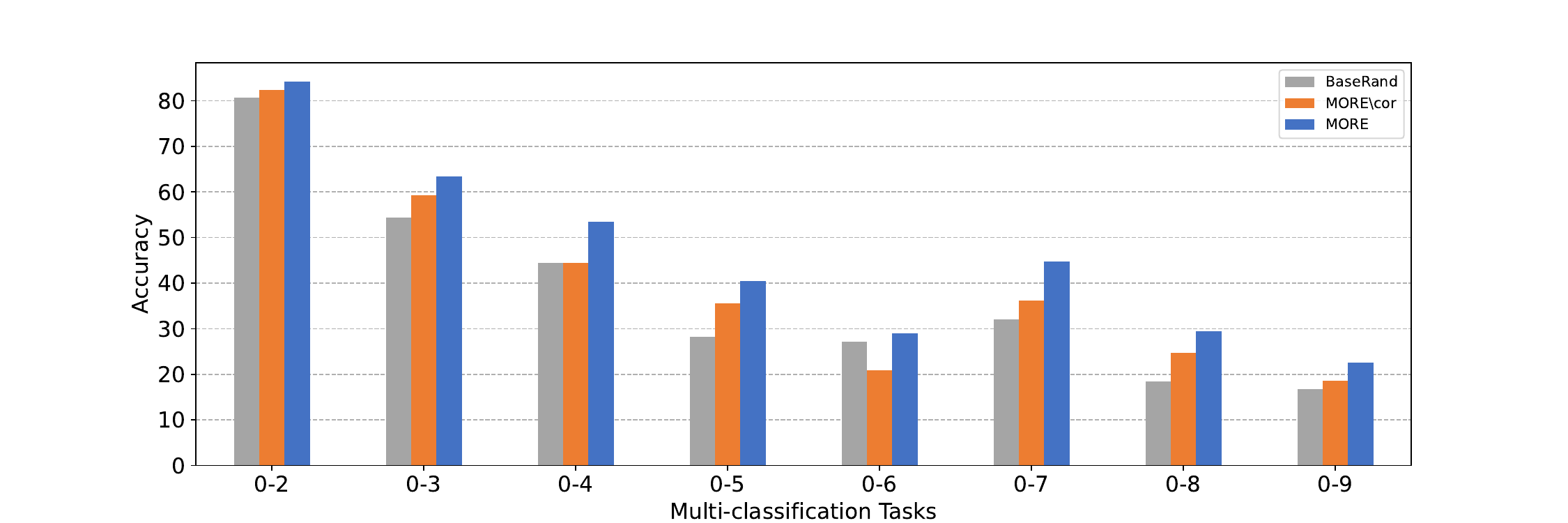}
\caption{Test accuracy of multi-classifications over various quantum label selection strategies}
\label{fig:eval-labels}
\end{figure*}

\subsection{Accuracy}
We evaluate the accuracy of the proposed approach \textbf{MORE} using the MNIST dataset on noise-free quantum systems. We conduct nine classification problems ranging from binary to 10-class. 

Our evaluation starts by investigating the effect of using multiple observables of a readout qubit on binary classification tasks. To do so, we compare our results to \textbf{BaseBin}, which serves as the baseline in this experiment. The comparison is illustrated in Fig.~\ref{fig:eval-bin}. Both MORE and BaseBin use the same ansatz (QCNN with 91 trainable parameters) for binary classifications, but the readout qubit of BaseBin is measured only with observable $\sigma_z$. The expected value of its measurement results is associated with two distinct classes: a positive expected value represents one class, and a negative expected value represents the other. We conduct 45 binary classifications on MNIST using MORE and BaseBin, respectively. The class pairs for classification are listed on the x-axis of Fig.~\ref{fig:eval-bin} and are sorted in descending order of interclass correlations. E.g., the training data of classes `4' and `9' have the highest similarity, while those of classes `0' and `1' have the lowest. The results indicate that MORE outperforms BaseBin in 37 out of 45 tasks and achieves comparable accuracy in the remaining tasks. MORE improves accuracy by up to 22.28\% and by an average of 4.9\%. Furthermore, the performance of MORE demonstrates greater stability across tasks of varying difficulty than BaseBin. As observed, a roughly inverse relationship exists between accuracy and interclass correlation for both MORE and BaseBin, i.e., as class correlation increases, classification becomes more challenging. Consequently, both MORE and BaseBin exhibit the highest accuracy on (0, 1)-classification and the lowest accuracy on (4, 9)-classification. Nonetheless, the variance of MORE's accuracy over 45 tasks is only 34.01, while that of BaseBin is 59.41, indicating that MORE is more stable than BaseBin. It shows the advantage of MORE, which has three observables, in terms of accuracy and stability.

Next, we evaluate MORE on multi-classification tasks using the MNIST dataset and summarize the results in Fig.~\ref{fig:eval-mul}. We conduct eight multi-classification tasks, categorizing handwritten digits into three to ten classes. Note that the ansatzes used in multi-classifications are \textit{identical} to the binary classification ansatzes. The accuracy of MOREs is compared to that of \textbf{BaseAnc} and \textbf{BaseMea}, which serve as the baselines in this experiment. BaseAnc and BaseMea are both variational quantum multi-classifiers. BaseAnc utilizes eight qubits for data processing and $n$ ancilla qubits as the readout for $n$-class classifications. And BaseMea employs a total of eight qubits and measures a subset of qubits at the end of the circuit to produce a result. In a multi-classification task with $n$ classes, $n$ qubits are measured. They encode class labels as one-hot vectors and measure the readout qubits on the z-basis. So BaseAn and BaseMea can only support the classifications involving up to eight classes, due to the limitation in dimensions and the number of qubits, respectively. MORE, however, is capable of conducting 10-class classification regardless of the number of qubits, and we believe it still qualifies if the dataset contains more classes. Fig.~\ref{fig:eval-mul} indicates that MORE outperforms BaseAnc and BaseMea across the board. The reason is that both BaseAnc and BaseMea require simultaneous control of several qubits to represent the class label, which can be challenging, particularly for an unstable quantum system. In contrast, MORE only utilizes a single qubit to record the outcome, thus reducing the number of factors contributing to instability. 

Overall, our proposed MORE approach, which uses a simple circuit with only one readout qubit, can beat other general approaches and achieve the desired performance. In the following subsections, we will analyze the impact of the MORE's components and access MORE in noisy quantum systems.


\subsection{Quantum label}

We now examine the impact of selecting quantum labels for classes. To do so, we compare the following approaches:
\begin{itemize}
    \item \textbf{BaseRand}: A baseline method that randomly assigns quantum states to classical labels.
    \item \textbf{MORE$\backslash$cor}: A variant of the MORE approach that does not consider interclass correlation during the variational quantum clustering step, i.e., all diagonal entries in the scaler array $S$ are -1s and all other entries are 1s.
    \item \textbf{MORE}: The vanilla MORE method employs interclass correlation for determining quantum labels.
\end{itemize}

Then, based on the selected quantum labels, each of the three approaches is followed by quantum label-based supervised learning using objective function Eq.~\ref{eq:sup_loss1} without the loss regulator $\mathcal{R}$. The test accuracy of the above approaches is summarized in Fig.~\ref{fig:eval-labels}. Across eight multi-classification tasks, MORE outperforms other methods in all cases. BaseRand and MORE$\backslash$cor have comparable performance, all of which are inferior to MORE. Therefore, we conclude that it is beneficial to take interclass correlations into account when deciding the quantum labels for classical labels. A possible reason is that the data distribution in the Hilbert space is strongly related to that in the classical feature space. As an example, Fig~\ref{fig:overview} (d) shows the distribution of quantum labels for class `0', `1', and `2' of MNIST, according to their relationship as listed in Fig~\ref{fig:overview} (b). Classes `1' and `2' have the smallest MSE value (strongest correlation), so their quantum labels are closer to each other compared to other class pairs. Similarly, the readout states of the instance from classes `1' and `2' are supposed to be closer than those of other classes. As a result, assigning quantum labels based on class correlation can capture the training data pattern to some extent, which is advantageous for enhancing model quality, as shown in Fig.~\ref{fig:eval-labels}. Nonetheless, the figure reveals that the accuracy of quantum classifiers decreases as the number of classes increases. This trend is also observed in quantum classifiers that employ alternative implementation strategies. Thus, we introduce a loss adjuster to alleviate this issue, as shown in Eq.~\ref{eq:reg}, and analyze its impact in the next subsection.

\subsection{Loss adjuster} \label{sec:reg}

\begin{table}[]
\centering
    \caption{Parameters and results for loss adjuster}
\begin{tabular}{cccccc} 
\toprule
Task & \begin{tabular}[c]{@{}c@{}}Min. label \\ distance\end{tabular} & $r$    & $w$   & More\textbackslash{}R acc. & MORE acc.\\ \midrule
0-2  & 1.302                                                          & 1.5  & 0.1 & 84.13\%                & 88.7\%  \\ \midrule
0-3  & 0.713                                                          & 1.0  & 0.2 & 63.45\%                  & 70.1\%  \\ \midrule
0-4  & 0.57                                                           & 0.8  & 0.1 & 53.5\%                   & 63.3\%  \\ \midrule
0-5  & 0.077                                                          & 0.2  & 0.4 & 40.48\%                  & 50.2\%  \\ \midrule
0-6  & 0.212                                                          & 0.4  & 0.5 & 29.06\%                  & 37.2\%  \\ \midrule
0-7  & 0.067                                                          & 0.2  & 0.5 & 44.8\%                   & 48.6\%  \\ \midrule
0-8  & 0.099                                                          & 0.15 & 0.2 & 29.4\%                   & 33\%    \\ \midrule
0-9  & 0.086                                                          & 0.15 & 0.2 & 22.6\%                   & 27.8\% \\
\bottomrule
\end{tabular}
\label{tab:lossreg}
\end{table}


Table~\ref{tab:lossreg} summarizes the parameters and results for the loss adjuster over the multi-classification tasks. The accuracy of \textbf{MORE} (with $\mathcal{R}$) and \textbf{MORE\textbackslash{}R} (without $\mathcal{R}$) are listed in the last two columns. The comparison indicates that MORE, which uses the loss function (Eq.~\ref{eq:loss_r}) with $\mathcal{R}$ during supervised learning, improves the accuracy across all tasks. This improvement is caused by the corrected misclassification between the classes whose quantum labels in the Hilbert space are too near together. In Eq.~\ref{eq:loss_r}, two hyper-parameters need to be determined by users: the threshold $r$ and the weight $w$. The threshold $r$ is empirically determined based on distances between quantum labels. In our experiments, the Cosine distance between quantum labels is calculated. Usually, $r$ is specified to be slightly larger than the shortest distance, as shown in the 2nd and 3rd columns of table~\ref{tab:lossreg}. E.g., the smallest Cosine distance between quantum labels in the 0-9 task, a 10-class classification, is 0.086 (about 24 degrees), so we set the $r$ to be 0.15 (about 32 degrees around the target quantum label). Moreover, we test MORE with a $w$ range from 0.1 to 1.0 and identify the optimal value of $w$ for each task. We report these values in the 4th column of the table, although we found accuracy improvements across all values of $w$ in each task. The $w$ values demonstrate that $\mathcal{R}$ has varying importance across different tasks, but its contribution is not greater than half in any task. Hence, we recommend that users search for an appropriate value of $w$ within the range of 0.1 to 0.5.

 

\subsection{Noisy quantum system}

\begin{table}[]
    \centering
    \caption{Test accuracy of MORE on Iris dataset with noisy backends}
    \begin{tabular}{cccc}
\toprule
\# Shots & FakeAuckland & FakeAthensV2 & FakeBelemV2 \\ \midrule
1k    & 95.56        & 95.56        & 93.33       \\ \midrule
2k    & 97.78        & 93.33        & 95.56       \\ \midrule
3k    & 97.78        & 93.33        & 95.56       \\ \midrule
4k    & 97.78        & 93.33        & 95.56       \\ \midrule
5k    & 93.33        & 93.33        & 95.56       \\ \midrule
6k    & 95.56        & 95.56        & 95.56       \\ \bottomrule
\end{tabular} 
    
    \label{tab:iris}
\end{table}

Table \ref{tab:iris} summarizes the test accuracy of MORE using the noisy quantum backends, including FakeAuckland, FakeAthensV2, and FakeBelemV2 on the Iris dataset. The hybrid quantum-classical training method of QNN typically requires a long time to execute on a quantum machine, but the IBM cloud quantum computing platform imposes time limits for tasks. In this experiment, we therefore utilize simulators of noisy quantum computing systems. The noisy backends used in this experiment have the noise model collected from real quantum machines, which includes T1 and T2 time, 1-qubit and 2-qubit gate errors, and measurement errors. So these simulated backends provide us with a practical and reasonable way to evaluate the performance of MORE in noisy quantum systems. To process the Iris instances of size 4, we construct 4-qubit QNNs with 39 trainable parameters. We update the QNNs for 300 steps using the COBYLA optimizer. We varied the number of shots for the quantum program from 1,000 to 6,000 on the backends to examine the impact of quantum computation costs.


The MOREs reach their highest accuracies of 97.78 (44/45), 95.56 (43/45), and 95.56 (43/45) on the FakeAuckland, FakeAthensV2, and FakeBelemV2 backends, respectively. The noisy MOREs achieve comparable performance to the noise-free version, with an accuracy of 97.78\%. Surprisingly, we found that the accuracy was not directly proportional to the number of shots, suggesting that the quantum computation cost required for this task is not significant. However, we acknowledge that the number of shots required may increase with an increase in the number of classes. Nonetheless, the cost is still acceptable only if the distribution of learned quantum labels is scattered in the Hilbert space. Furthermore, while it is true that accurate measurement results are necessary for the MORE approach, we want to highlight that using a simple quantum circuit (with fewer qubits, gates, and shallow depth) can lead to less error compared to other methods. This makes the MORE approach feasible and promising to implement on NISQ machines.

\section{Related work} \label{sec:related}
In recent years, numerous attempts have been made to explore the capabilities of QNNs using various techniques. In this section, we will discuss some notable works that focus primarily on the classification task, which is the central theme of this article.

First of all, there are several suggested concepts for the design of QNN-based classifiers \cite{farhi2018classification, cong2019quantum, henderson2020quanvolutional, wu2022wpscalable}. \cite{farhi2018classification} and \cite{cong2019quantum} concentrate on the ansatz design of QNNs. They are the fundamentals of current popular QNNs that use parametric unitary transformations to process data efficiently. However, scalability is a crucial concern for QNNs, particularly on NISQ devices, where the limited number of qubits restricts the quantum circuit's size and performance. Quanvolutional NN \cite{henderson2020quanvolutional} employs the quantum circuit of modest size as kernels of quantum convolution NN, rather than building the entire QNN. And SQNN \cite{wu2022wpscalable} scales up QNNs by constructing large-scale QNNs modularly and leveraging quantum computation resources from multiple quantum machines. All of these studies focus on binary classification tasks and demonstrate the advantages of QNNs. In addition to model scalability, the ability to scale the problem size is essential for improving the capacity of QNNs. In this article, we propose an approach that aims to expand the problem size that a QNN can handle without having to scale up its quantum circuit.

Many prior efforts have attempted to enhance the problem size that QNN classifiers can address \cite{bokhan2022multiclass, li2021vsql, nghiem2021unified, wu2020end, blank2020quantum, schuld2017implementing, yun2022projection, chalumuri2021hybrid, yang2020entanglement}. A QNN classifier with four ancilla qubits for 4-class classification is built in \cite{bokhan2022multiclass}. The capability of the solution is restricted by the quantum hardware, as larger problem sizes require additional ancilla qubits, which are not currently available on NISQ machines. \cite{chalumuri2021hybrid, yang2020entanglement} also address milt-classification problems by using ancilla qubits. An alternative solution is QuClassi proposed in \cite{stein2022quclassi}. It breaks down a multi-classification into several binary classifications. For each binary classification, the cost function is constructed based on quantum state fidelity by using SWAP, and only one ancilla qubit is needed for measurement to get the readout. The measurement results of all classes are softmaxed to obtain the final classification result. Despite the quantum computing resources being limited in this method, the training and inference processes are complicated and time-consuming. Similarly, a fidelity-based QNN classifier is proposed in \cite{schuld2017implementing}, but its QNN ansatz can only solve binary classification tasks. Moreover, VSQL \cite{li2021vsql} slides a predefined partial observable ``xx" across the qubits to get the classical shadows of the quantum state prepared by a QNN, and then feeds the local shadows to a classical fully-connected layer for decision making. Yet, the computationally demanding classical fully-connected layer may counteract the quantum speedup. A unified quantum classifier is presented in \cite{nghiem2021unified}. This approach investigates the connection between class labels and quantum states. In our study, we further investigate the distribution of quantum labels in Hilbert space based on interclass correlations and demonstrate its significance by highlighting the superior performance of our approach. Additionally, we introduce a loss adjuster during supervised learning to ensure optimal performance when dealing with multiple classes in the task at hand.




\section{Conclusion} \label{sec:conclusion}

We propose MORE, a QNN-based multi-classifier that maximizes quantum resource efficiency. By fully leveraging quantum information, MORE employs a simple ansatz as the binary classifier with only one readout qubit to solve multi-classification problems. To achieve this, MORE first converts classical labels into corresponding quantum labels using variational quantum clustering. This process considers interclass correlations, allowing the learned quantum labels to capture intrinsic patterns in the classical feature space of the training data. Quantum supervised learning is then performed based on the quantum labels to efficiently train the model. Furthermore, we introduce a loss adjuster to enhance the model's quality by making it more sensitive to the labels that can result in misclassifications during training. Our comprehensive evaluations demonstrate that MORE outperforms general quantum multi-classifiers and is capable of effective operation in noisy quantum systems due to its reduced error sources. This suggests that fully utilizing quantum information offers a promising approach for scaling up the problem sizes that can be addressed during the current NISQ era.

\section*{Acknowledgment}
The authors would like to thank all the reviewers for their helpful comments. 
This project was supported in part by US National Science Foundation grant CNS-1816399. This work was also supported in part by the Commonwealth Cyber Initiative, an investment in the advancement of cyber R\&D, innovation and workforce development. For more information about CCI, visit cyberinitiative.org.

\bibliographystyle{IEEEtran}
\bibliography{ref}

\end{document}